\documentclass[12pt]{article}
\usepackage{latexsym, amsmath}
\begin{document}
\begin{center}
{\bf ANOMALY INDUCED EFFECTIVE ACTION COMBINED WITH
GENERALIZED BRANS-DICKE THEORY }\\

\bigskip

\bigskip

Iver Brevik \\

\bigskip

Department of Energy and Process Engineering, Norwegian University
of Science and Technology, N-7491 Trondheim, Norway.

Email: iver.h.brevik@ntnu.no

\bigskip

\begin{abstract}

We consider the anomaly induced effective action in $\cal{N}$ =4
super Yang-Mills theory in interaction with the Brans-Dicke (BD)
field. The generalization of the BD theory so as to permit an
energy exchange between the scalar field and ordinary matter
fields, was recently worked out by T. Clifton and J. D. Barrow
[Phys. Rev. D {\bf 73}, 104022 (2006)]. We derive the scalar field
equations for the dilaton field, and the BD field, and discuss the
Friedmann equation in the general case. The present paper is a
continuation of an  investigation some years ago dealing with the
case of conformal anomaly plus ordinary classical gravity [I.
Brevik and S. D. Odintsov, Phys. Lett. B {\bf 455}, 104 (1999)].

\end{abstract}
\end{center}

\bigskip


\section{Introduction}

The recent paper of Clifton and Barrow \cite{clifton06} sheds
light on an interesting possibility in the formulation of
scalar-tensor gravity theories, namely to allow for an {\it
exchange} of energy (and momentum) between the scalar field $\phi$
and the ordinary matter fields. This kind of generalization
implies an enrichment of the formalism; as these authors point
out, one may even obtain variations in the gravitational
"constant" $G$. the field-matter exchange means that the
energy-momentum tensor $T_M^{\mu\nu}$ of matter is no longer
divergence-free,
\begin{equation}
\nabla_\nu T_M^{\mu\nu}=f^\mu, \label{1}
\end{equation}
where $f^\mu$ is the force density from the field on the matter.
What we shall focus attention  in the following, is the
Brans-Dicke theory \cite{brans61}.

First of all we note, however, that this separation into two
interacting  subsystems - fields and matter - bears a striking
resemblance to the theory of the electromagnetic field in
continuous media. Actually, this is the key element in the famous
Abraham-Minkowski controversy, a problem that has been discussed
with more or less intensity since Abraham and Minkowski formulated
their energy-momentum expressions around 1910. The advent of
accurate experiments, in particular, has helped us to get more
insight into this complicated field-matter interacting system.
Some years ago, the present author  made a review of the
experimental status in the field \cite{brevik79}. There is by now
a rather big literature in this field; some papers are listed in
Ref.~\cite{moller72,kentwell87,antoci98,obukhov03,loudon97,
garrison04,feigel04,leonhardt06}.

What we shall consider in the following is however a different
theme, namely to what extent the inclusion of the {\it conformal
anomaly} for scalar fields influences the equations of
 the interacting Brans-Dicke theory. We shall consider quantum
 ${\mathcal N}=4$ super Yang-Mills theory interacting in covariant way
 with  ${\mathcal N}=4$ conformal supergravity. The induced large $N$
 effective action for such a theory can be calculated on a
 non-supersymmetric dilatonic-gravitational background using the
 conformal anomaly found via the AdS/CFT correspondence. Our
 intention is to show this calculation in the Brans-Dicke
 background. On a purely bosonic background with only non-zero
 gravitational and dilaton fields, the conformal anomaly for
 ${\mathcal N}=4$ super YM theory was calculated in
 \cite{nojiri98} via the AdS/CFT correspondence
 \cite{maldacena98,witten98,gubser98}. Our calculation is a
 generalization of that presented earlier in Ref.~\cite{brevik99}.

 We mention that four-dimensional quantum cosmology with account
 taken of dilaton-dependent conformal anomaly was considered in
 Refs.~\cite{gates98,nieuwenhuizen99,nojiri01}.

 We consider only the quantum effects of the ${\cal N}=4$ super
 Yang-Mills theory together with the Brans-Dicke theory as a
 specific example. We might in principle consider an arbitrary
 conformal quantum field coupled with a dilaton different from the
 Brans-Dicke scalar. This would amount to different numerical
 values for the coefficients of conformal anomaly, and in some
 cases it would lead to the appearance of new, dilaton dependent
 terms \cite{nojiri01}.

 In the next section we establish the field equations for the
 dilaton field $\chi$ and the Brans-Dicke field $\phi$, and find
 the Friedmann equation under specified simplifying conditions. In
 section 3, some typical properties of the solutions of the
 Friedmann equation are discussed.

 \section{Basic formalism}

 Let us start from the anomaly induced effective action
 \cite{reigert84}. We write it in a non-covariant local form, and
 we limit ourselves to a conformally flat metric, i.e.,
 $g_{\mu\nu}=e^{2\sigma} \eta_{\mu\nu}$, where $\eta_{\mu\nu}$ is
 the Minkowski metric. We assume also that only the real part of
 the dilaton field $\chi$ is different from zero.

 The anomaly induced effective action takes under these conditions a relatively
 simple form:

 \[ W=-\int d^4 x \Big\{ 2b' \sigma \Box^2 \sigma-3[b''
 +\frac{2}{3}(b+b')] \]
 \begin{equation}
 \times (\Box \sigma +\partial_\mu \sigma \partial^\mu
 \sigma)^2+C\sigma \chi \, \Box^2 \chi \Big\}. \label{2}
 \end{equation}
 Here $\Box =\partial^\mu \partial_\mu$, all derivatives being
 flat ones. The various constants are given as
 \begin{equation}
 b=\frac{N^2-1}{(4\pi)^2}\frac{N_s+6N_f+12N_v}{120}, \label{3}
 \end{equation}
 \begin{equation}
 b'=-\frac{N^2-1}{(4\pi)^2}\frac{N_s+11N_f+62N_v}{360}, \label{4}
 \end{equation}
 \begin{equation}
 C=\frac{N^2-1}{(4\pi)^2}N_v. \label{5}
 \end{equation}
 In the ${\cal N}=4$ SU(N) super YM theory, the scalar, fermion
 and vector numbers are $N_s=6, N_f=2, N_v=1$, giving
 \begin{equation}
 b=-b'= \frac{C}{4}=\frac{N^2-1}{4(4\pi)^2}. \label{6}
 \end{equation}
 We  assume that the scale factor $a(\eta)$ depends only on
 conformal time, $\sigma(\eta)=\ln a(\eta)$. Also, $\chi$ is assumed to depend only on $\eta$. One now has to add to
 the action (\ref{2}) the contribution from the Brans-Dicke field,  and from matter:
 \begin{equation}
 S_{BD}=\frac{1}{16\pi}\int d^4 x \sqrt{-g}\left( \phi
 R-\frac{\omega}{\phi}\partial_\mu \phi \partial^\mu \phi +16\pi L_M
 \right). \label{7}
 \end{equation}
 Here $R$ is the curvature scalar, $\omega$ is the Brans-Dicke
 coupling constant, and $L_M$ is the Lagrangian density of the
 matter fields. If $\omega \rightarrow \infty$ this theory reduces
 to general relativity. Actually, by putting $1/\omega =G$ we
 recover straightaway the action integral for classical general relativity
 \begin{equation}
 S_{cl}=\frac{1}{16\pi G}\int d^4 x \sqrt{-g}\, (R+16\pi G L_M). \label{8}
 \end{equation}
 (Note that in Ref.~\cite{brevik99}, $W$ and $S_{cl}$ were defined
 with the opposite sign; our present convention is made to agree with  the
 standard convention  for the Brans-Dicke term (\ref{7}).)

 There are thus two scalar fields present: the field $\chi$ coming
 from the conformal anomaly, and the Brans-Dicke field $\phi$.
 Note that $\chi$ is defined to be dimensionless, while the
 dimension of $\phi$ is ${\rm cm}^{-2}$.
 These fields may be pictured as two "fluids". Also, the matter
 fields may be associated with a material fluid.

 Consider first the field equation for $\chi$. Since the
 Brans-Dicke action does not depend on $\chi$ we get the same
 equation as in the conformal anomaly-gravity case \cite{brevik99}:
 \begin{equation}
 \ln a\, \chi'''' +(\ln a\, \chi)''''=0, \label{9}
 \end{equation}
 prime meaning differentiation with respect to $\eta$. We can here
 make a transformation to cosmological time $t$ via the relation
 $dt=a(\eta)d\eta$. The result is rather complicated,
 \[ 2a \ln a\,Y[\chi,a]+\chi a^3 \stackrel{....}{a} a  +4a^3\dot{a}\stackrel{...}{a}\chi+
 3a^2\chi \, \dot{a}\stackrel{...}{a} \]
 \[+4a^3\dot{\chi}\stackrel{...}{a}+ 6a^3\ddot{\chi}\ddot{a}+12a^2\ddot{\chi}\ddot{a}+12a^2\ddot{\chi}\dot{a}^2
 +14a^2\dot{\chi}\,\dot{a}\ddot{a} \]\begin{equation}
 +a\chi \ddot{a}\dot{a}^2 +a^2\chi
 \ddot{a}^2+4a\dot{\chi}\,\dot{a}^3=0, \label{10}
 \end{equation}
 where
 \[ Y[\chi,a]=a^3 \stackrel{....}{a}\chi
 +6a^2\dot{a}\stackrel{...}{a}\chi+4a^2\ddot{a}\ddot{\chi}+7a\dot{a}^2\ddot{\chi}\]
 \begin{equation}
 +4a\dot{a}\ddot{a}\dot{\chi}+a^2\stackrel{...}{a}a\dot{\chi}+\dot{a}^3\dot{\chi}.
 \label{11}
 \end{equation}
 We shall consider only approximate solutions of Eq.~(10) when the
 term with $\ln a$ can be dropped.

 The field equation for $\phi$ can be found analogously. Since $W
 $ does not depend on $\phi$ we have to vary the expression
 (\ref{7}) only. We assume that the material fluid has energy
 density $\rho$ and pressure $p$, and assume that the equation of
 state can be written in the conventional form
 \begin{equation}
 p=(\gamma-1)\rho, \label{12}
 \end{equation}
 where $\gamma$ is a constant, the case $\gamma=2$ (Zeldovich
 fluid) denoting the upper limit for $\gamma$. The field equation
 for $\phi$ becomes the same as in the pure Brans-Dicke case
 \cite{clifton06}:
 \begin{equation}
 \ddot{\phi}+3H\dot{\phi}=\alpha (4-3\gamma)\rho -2\alpha
 \frac{\phi}{\dot{\phi}} \,f^0, \label{13}
 \end{equation}
 where
 \begin{equation}
 \alpha=\frac{8\pi}{3+2\omega}. \label{14}
 \end{equation}
 Finally there is the Friedmann equation, obtained by varying the
 total action $W+S_{BD}$ with respect to the metric tensor. In
 order to simplify the formalism somewhat, we exploit that the
 coefficient $b''$ is known to be ambiguous. We have a freedom in
 choosing its value. As advocated in Ref.~\cite{liu98}, one can
 well take
 \begin{equation}
 3b''+2b=0. \label{15}
 \end{equation}
 (The physical advantage of this choice is that no $\Box R$-term
 remains in the conformal anomaly.) As for the conformal anomaly
 part, considerable simplification of the Friedmann equation
 thereby occurs. In the pure conformal anomaly case, we can write
 the square of the Hubble parameter as \cite{brevik99}
 \begin{equation}
 H^2\Big|_{CA}=-\frac{1}{24}\frac{C\chi Y[\chi,
 a]}{b'\,\ddot{a}a^2}. \label{16}
 \end{equation}
 (Here the contribution from the $R$-term $S_{cl}$ in the action, Eq.~(6) in \cite{brevik99}, has been
 omitted.) In the
 Brans-Dicke case \cite{clifton06}
 \begin{equation}
 H^2 \Big|_{BD}=\frac{8\pi}{3\phi}\rho
 +\frac{\omega}{6}\frac{\dot{\phi}^2}{\phi^2}-H\frac{\dot{\phi}}{\phi},
 \label{17}
 \end{equation}
 leading to the conventional relation $H^2=(8\pi G/3)\rho$ in the
 GR case ($\phi=1/G$).
 Now combining expressions (\ref{16}) and (\ref{17}) we may
 construct
 the Hubble parameter in the general case as
 \[ H^2=-\frac{1}{24}\frac{C\chi \, Y[\chi,
 a]}{b'\,\ddot{a}a^2}\]
 \begin{equation}
+\frac{8\pi}{3\phi}\rho
 +\frac{\omega}{6}\frac{\dot{\phi}^2}{\phi^2}-H\frac{\dot{\phi}}{\phi}.
 \label{18}
 \end{equation}
 Obviously, this expression agrees with known expressions in the
 two limiting cases.

 \section{Remarks on the solutions }

 We will not discuss in detail the solutions of the complicated
 equation (\ref{18}), but will note some general
 properties that can be verified almost by a mere inspection of the
 equation.

 First of all, one may ask at which stage in the history of the
 universe the influence from the first term in Eq.~(\ref{18})
 was important. In order to investigate this point, we may replace
 the
 Brans-Dicke theory with general relativity again and estimate the ratio between
 the magnitudes of the conformal anomaly
 term
 and  the classical term, $(8\pi /3\phi)\rho \rightarrow (8\pi
 G/3)\rho$. Assuming radiation dominance in the very early
 universe, meaning in standard notation that
 $\rho=\rho_0(a_0/a)^4$, we have
 \begin{equation}
 a=\left(\frac{32\pi G}{3}\rho_0a_0^4\right)^{1/4} t^{1/2},
 \label{19}
 \end{equation}
 leading to $\rho=3/(32\pi Gt^2)$. From Eq.~(\ref{6}), it follows
 that $-b' \sim C$. As for the term $Y[\chi, a]$, we let its
 magnitude be represented by the first term on the right in
 Eq.~(\ref{11}). Making use of Eq.~(\ref{19}) we then obtain after
 a brief calculation that the ratio between the first and the
 second term to the right in Eq.~(\ref{18}) becomes of order
 $a\chi^2$, thus quite simple. Now, on physical grounds we would
  expect $\chi$ to decay exponentially with time. Let us express
  this as
 $\chi \propto e^{-\beta \tilde{H}t}$, to conform with the notation in Eq.~(\ref{20}) below.  The expression
 $a\chi^2$ thus approaches zero when $t\rightarrow 0$. At the
 beginning of the universe, the relative importance of conformal anomaly was equal to zero.
 The influence of conformal anomaly decreases to zero for large times also, like $t^{-1/2}
 e^{-2\beta \tilde{H}t}$. The maximum contribution occurs at an intermediate time, $t=1/(2\beta \tilde{H})$.

 We now consider two special examples.

 \subsection{The case of conformal anomaly-general relativity}

 This case was discussed in an earlier paper \cite{brevik99}. For
 completeness, we summarize the main results also here.

 We look for approximate solutions based upon the de Sitter forms:
 \begin{equation}
 a(t)=\tilde{a}_0e^{\tilde{H}t}, \quad \chi(t)=\chi_0 e^{-\beta
 \tilde{H}t}, \label{20}
 \end{equation}
 where $\tilde{H}$ and $\beta$ are constants. We consider only very
 early times, for which $\tilde{H}t \ll 1$. It implies, as
 mentioned above, that the $\ln a$ term in Eq.~(\ref{10}) can be
 dropped. From the remaining part of Eq.~(\ref{10}) we obtain
 three different modes, corresponding to $\beta= \{(3/2), 2.62,
 0.38 \}$. For each  mode we find
 \begin{equation}
 \tilde{H}^2=-\frac{1}{16\pi G}\left[ b'+\frac{C}{24}\chi_0^2\,
 (\beta^4-6\beta^3+11\beta^2-6\beta) \right]^{-1}. \label{21}
 \end{equation}
 For the modes $\beta= 2.62$ or $\beta= 0.38$, we calculate $\tilde{H}$ to
  be less than the Hubble parameter in the non-dilatonic case. The
  effect of the dilaton is thus to slow down the inflation.

  For the mode $\beta=3/2$ the situation is reversed: the value of
  $\tilde{H}$ is larger than in the non-dilatonic case and the
  dilaton acts to speed up the inflation.

  \subsection{The case $C=0$ }

  We now assume the general conformal anomaly - Brans Dicke case,
  but impose the condition that the coefficient $C$ is equal to
  zero. According to Eq.~(\ref{5}) this means that $N_v=0$. The
  vector contribution to the super YM theory is thus omitted.

  The effect that this simple condition has on the formalism is
  dramatic: in the Friedmann equation (\ref{18}) there is no
  influence from the conformal anomaly at all. The Hubble
  parameter evolves in time as if the Brans-Dicke field were the
  only scalar field. This does not imply that the dilaton field is
  equal to zero, however. Still, $\chi$ evolves according to the field
  equations (\ref{9}) or (\ref{10}).

  Similarly, the condition $C=0$ is seen to be important also in
  the expression (\ref{21}). We now get simply $\tilde{H}^2=-1/(16\pi
  G b')$, so that there is no dilatonic influence on the Hubble
  parameter.

  \section{Summary}

  The starting point for our analysis of the situation where there
  is both a conformal anomaly dilaton field $\chi$ and a
  Brans-Dicke field $\phi$ present, was to combine the actions in
  Eqs.~(\ref{2}) and (\ref{7}). We found the field equation for
  $\chi$ to be  given by Eq.~(\ref{9}) (in terms of conformal time
  $\eta$), or by Eq.~(\ref{10}) (in terms of cosmological time
  $t$). The field equation for $\phi$ is given by Eq.~(\ref{13}).
  This equation is written in a form general enough to permit
  interaction between $\phi$ and the matter fields
  \cite{clifton06}.

  A main point in the analysis was to construct the expression for
  $H^2$ in Eq.~(\ref{18}) as the sum of the
  expressions (\ref{16}) and (\ref{17}) for the two separate
  subsystems. This form for the Friedmann equation is based upon
  the choice of the gauge condition (\ref{15}), which is a
  permissible choice because of the ambiguity of the coefficient $b''$.
  We found the relative importance of the conformal anomaly to go
  to zero both for $t\rightarrow 0$ and for large values of $t$,
  and to have a maximum at some intermediate time.

  The special case $C=0$ is of interest. It corresponds to
  omitting the
  vector contribution ($N_v=0$) to the super YM theory. Then, the influence
  from conformal anomaly on the Hubble parameter becomes simply
  zero. We get  the same solutions for the Brans-Dicke field as
  if there were no conformal anomaly at all, even under the very
   general conditions studied in \cite{clifton06}.

   In general, we have discussed the situation where the scalar field from the
   conformal anomaly is different from the scalar field in the
   Brans-Dicke theory (i.e., ${\cal N}=4$ SYM is considered in the
   Brans-Dicke background). However, one can also identify the
   dilaton from conformal anomaly with the Brans-Dicke scalar, as
   was done in Ref.~\cite{geyer99}. Moreover, other backgrounds
   can also be considered, like the AdS background
   \cite{brevik00}.

   Finally, it is of interest to note that our problem may easily
   be generalized to the case of a brane-world scenario with a
   dilaton dependent conformal anomaly and inflation
   \cite{nojiri00,nojiri00a,nojiri00b}.

\subsection*{Acknowledgment}

I thank Sergei D. Odintsov for his many valuable comments on the
manuscript.

\end{document}